\documentclass[11pt]{article}

\begin{document}

\setlength{\textwidth}{130mm} \setlength{\textheight}{194mm}

\title{Twistorial and space-time descriptions \\
 of $D=4$ string models\thanks{Presented at XXII Max Born Symposium (September 2006, Wroc{\l}aw by
 S. Fedoruk)}}

\author{Sergey  Fedoruk\thanks{BLTP JINR,
Dubna, Russia  \tt{e-mail: fedoruk@theor.jinr.ru}}
,$\;$ Jerzy Lukierski\thanks{ITP
University of Wroc{\l}aw, Poland \tt{e-mail: lukier@ift.uni.wroc.pl}}
}
\date{}
\maketitle

\begin{abstract}
We derive twistorial tensionful bosonic string action by considering
 on the world sheet
  the canonical twistorial
2--form in two--twistor space.
We demonstrate the
 equivalence of or model  to two
 known momentum formulations of $D=4$
  bosonic string,
   with covariant worldsheet vectorial string momenta
$P_\mu^{\,m}(\tau,\sigma)$ and the one with tensorial string momenta
$P_{[\mu\nu]}(\tau,\sigma)$. All considered here string actions, in twistorial and mixed
spinor-spacetime formulations, are classically  equivalent
 to the Nambu-Goto action.
\end{abstract}

\section{Introduction}

Twistors and supertwistors~\cite{PenMac,Fer} have been often  used
for the description of (super) particles and (super) strings, as
an alternative to space--time approach (see {\it
e.g.}~\cite{Shir}-\cite{Sor}). Recently large class of
perturbative amplitudes in $N=4$ $D=4$ supersymmetric Yang-Mills
theory \cite{Wit} and conformal supergravity (see e.g.
\cite{BerWit}) were described in a simple way by using strings
moving in supertwistor space.

Our main aim of this report which is based on our paper~\cite{FL} is to derive the
twistorial master action is classically equivalent to
$D=4$ Nambu-Goto tensionful string model.

We present here
  the twistorial formulation of tensionful string
    in
contrast
 with other studies which provide only the  twistorial
  null (super)string and
 twistorial
   massless (super)particle models.
 In this paper we shall consider
 pure space--time
formulation, pure twistor description and intermediate mixed
space-time--twistorial
formulation.

In order to relate space-time and twistor target geometries
 for the formulation of the classical theory we shall
employ the Penrose incidence relations.

Twistorial formulation of massive particles with spin~\cite{FedZim,azbelu,azfrlu,FFLM} in
$D=4$ space--time is described in two--twistor space. The corresponding action is
constructed from the following two--twistor one--form ($A=1,...,4$ is the $SU(2,2)$ index)
\begin{equation}\label{1-form-x-i}
\Theta^{(1)} = \sum_{i=1}^2\Theta^{(1)}_i = \sum_{i=1}^2
\left(\bar Z^{Ai} dZ_{Ai}- d\bar Z^{Ai} Z_{Ai}\right)
\end{equation}
with imposed suitable constraints.

In this report we show that the
 action determining the  twistor formulation
of the tensionful
  string is defined in two--twistor space via embedding of
    the canonical Liouville two--form in
two--twistor space
\begin{equation}\label{2-form}
\Theta^{(2)} =  \Theta^{(1)}_1 \wedge \Theta^{(1)}_2\,.
\end{equation}
As we shall demonstrate, this twistor formulation is
 classically equivalent to the Hamiltonian formulation of
$D=4$ bosonic free string theory both with vectorial and tensorial
string momenta.

\section{Space--time formulations of the tensionful string}

The tensionful string in flat Minkowski space is described by
nonlinear Nambu--Goto action \footnote{The indices $m,n=0,1$ are
vector world-sheet indices; $\mu,\nu=0,1,2,3$ is vector
space--time ones.}~\cite{Nambu,Goto}
\begin{equation}\label{action-NG}
S=-T\int d\,^2\xi\, \sqrt{-\det (g_{mn})}
\end{equation}
where $\xi^m=(\tau,\sigma)$ are world--sheet coordinates,
\begin{equation}\label{ind-metr}
g_{mn}= \partial_m X^\mu\partial_n X_\mu
\end{equation}
is the induced metric in string surface, $T$ is the string tension.

Generally, the transition to twistorial formulation is
 realized via
 `Ha\-mil\-tonian' formulations.
 In case of the tensionful string~(\ref{action-NG})
there are two
 Hamiltonian frameworks:

{\bf Formulation with vectorial momenta.} The first order
formulation of the tensionful string~(\ref{action-NG}) is defined
by the action~\cite{Sieg2}
\begin{equation}\label{act-s}
S= \int  d\,^2\xi \left[P_\mu^{\,m}\partial_m X^\mu + {\textstyle
\frac{1}{2T}}(-h)^{-1/2}h_{mn}P_\mu^{\,m} P^{\,\mu\,n}\right]
\end{equation}
used vectorial momenta variables $P_\mu^{\,m}(\xi)$. The kinetic
part of the action~(\ref{act-s}) is described by the two--form
\begin{equation}\label{Th-2}
\tilde\Theta^{(2)} = P_\mu \wedge d X^\mu
\end{equation}
where $P_\mu =P_\mu^{\,m}\epsilon_{mn} d \xi^n$, $d X^\mu = d
\xi^m \partial_m X^\mu$ {\it i.~e.} in the
formulation~(\ref{act-s}) the pair $(P_\mu^{\,0},P_\mu^{\,1})$ of
generalized string momenta are represented by a one--form.

The equations of motion for world--sheet metric $h_{mn}$
\begin{equation}\label{3}
P^{\,m}_\mu P^{\,n \mu}-{\textstyle \frac{1}{2}}h^{mn}h_{kl}P^{\,k}_\mu P^{\,l \mu}=0
\end{equation}
describe the Virasoro first class constraints.

Expressing $P^{\,m}_\mu$ in the action~(\ref{act-s}) by its
equation of motion
\begin{equation}\label{2}
P^{\,m}_\mu= -T(-h)^{1/2}h^{mn}\partial_n X_\mu
\end{equation}
one obtains second-order action~\cite{Pol}
\begin{equation}\label{action-2}
S=-{\textstyle \frac{T}{2}}\int d^2\xi (-h)^{1/2}h^{mn}\partial_m X^\mu\partial_n X_\mu\,.
\end{equation}

{\bf Formulation with tensorial momenta.} Other formulation of the
bosonic string~(\ref{action-NG}) is the model with tensorial
momenta. It is obtained from the Liouville two--form
\begin{equation}\label{Th-2-2}
\tilde{\tilde\Theta}{}^{(2)} = P_{\mu\nu} d X^\mu \wedge d
X^\nu\,.
\end{equation}
Such a model is directly related with the interpretation of
strings as dynamical world sheets with the surface elements
\begin{equation}\label{els}
d S^{\mu\nu} = d X^\mu \wedge d X^\nu =
\partial_m X^\mu \partial_n X^\nu \epsilon^{mn} d^2\xi \,.
\end{equation}
The string action with tensorial momenta is
\begin{equation}\label{tens-4}
S = {\textstyle\sqrt{2}} \int  d\,^2\xi  \left[ P_{\mu\nu}\,
\partial_m  X^{\mu} \partial_n  X^{\nu} \epsilon^{mn}  -
\Lambda\left(P^{\mu\nu} P_{\mu\nu}  +  {\textstyle\frac{T^2}{4}}\right)  \right] .
\end{equation}

Expressing $P_{\mu\nu}$ by its equation of motion, we get
\begin{equation}\label{p-mn}
P^{\mu\nu}={\textstyle\frac{1}{2\Lambda}}\Pi^{\mu\nu}\,,  \qquad
\Pi^{\mu\nu}\equiv\epsilon^{mn}\partial_m X^{\mu} \partial_n X^{\nu}\,.
\end{equation}
It is important that the solution~(\ref{p-mn})  satisfies the
constraint $P^{\mu\nu}\tilde P_{\mu\nu}=0$ as an identity where
$\tilde P_{\mu\nu}=
\frac{1}{2}\epsilon_{\mu\nu\lambda\rho}P^{\lambda\rho}$. After
substituting~(\ref{p-mn}) in the action~(\ref{tens-4}) we obtain
the four--order action (see {\it e.g.}~\cite{BDM})
\begin{equation}\label{tens-5}
S={\textstyle\frac{1}{2\sqrt{2}}}\int d\,^2\xi
 \left[ \Lambda^{-1} \Pi^{\mu\nu}\Pi_{\mu\nu}
- \Lambda T^2 \right]\,.
\end{equation}
Eliminating further $\Lambda$ and using that
\begin{equation}
\Pi^{\mu\nu}\Pi_{\mu\nu} =2 \det (g_{mn})
\end{equation}
we obtain the Nambu--Goto string action~(\ref{action-NG}).

\section{Tensionful string in mixed twistor-spacetime formulation}

{\bf Mixed formulations with vectorial momenta.} In order to obtain from the
action~(\ref{act-s}) the mixed spinor--space-time action~(\ref{SSTV})
 we should eliminate
the fourmomenta $P_\mu^m$ by means of the the string generalization
of the Cartan--Penrose
formula. On curved world sheet it has the
form\footnote{$h_{mn}=e_m^a e_{n a}$ is a
world--sheet metric, $e_m^a$ is the zweibein, $e_m^a e^m_b = \delta^a_b$,
$e=\det(e_m^a)=\sqrt{-h}$. The indices $a,b=0,1$ are $d=2$ flat indices.
The indices
$i,j=1,2$ are $d=2$ Dirac spinor indices.
We use bar for complex conjugate quantities,
$\bar{\lambda}_{\dot\alpha}^i =(\overline{\lambda_{\alpha i}})$,
and tilde for Dirac
conjugated $d=2$ spinors, $\tilde{\lambda}_{\dot\alpha}^i =
\bar{\lambda}_{\dot\alpha}^j(\rho^0)_j{}^i$.}~\cite{FL}
\begin{equation}\label{P-res-st}
P_{\alpha\dot\alpha}^{\,m}=e\,
\tilde{\lambda}_{\dot\alpha}\rho^m\!\lambda_{\alpha} =e
e^m_a\tilde{\lambda}_{\dot\alpha}^i(\rho^a)_i{}^j\lambda_{\alpha
j}\,.
\end{equation}
Then the second term in string action~(\ref{act-s}) takes the form
\begin{equation}\label{2-term}
{\textstyle \frac{1}{2T}}(-h)^{-1/2}h_{mn}P_\mu^{\,m} P^{\,n \mu}={\textstyle
\frac{1}{2T}}\,e\, (\lambda^{\alpha i}\lambda_{\alpha i})
(\tilde{\lambda}_{\dot\alpha}^j\tilde{\lambda}^{\dot\alpha}_j)
\end{equation}
where we used ${\rm Tr}(\rho^m \rho^n) =2h^{mn}$. Putting~(\ref{P-res-st}) and
(\ref{2-term}) in the action~(\ref{act-s}) we obtain the string action~\cite{SSTV}
\begin{equation}\label{SSTV}
S=\!\int\!\! d^2\xi \, e\left[\tilde{\lambda}_{\dot\alpha}
\rho^m\!\lambda_{\alpha }\,
\partial_m X^{\dot\alpha\alpha}+ {\textstyle \frac{1}{2T}}\,(\lambda^{\alpha i}
\lambda_{\alpha i})
(\tilde{\lambda}_{\dot\alpha}^j\tilde{\lambda}^{\dot\alpha}_j)\right]
\end{equation}
which provides the mixed space-time--twistor formulation of bosonic string. We stress that
in the formulation~\cite{SSTV} the twistor spinors $\lambda_{\alpha i}$ are not
constrained. Further, the algebraic field equation~(\ref{3}) after
substitution~(\ref{P-res-st}) is satisfied as an identity.

The action~(\ref{SSTV}) invariants under the following local transformations:
$$
\lambda^\prime_{\alpha i} =e^{i(b+ic)}\lambda_{\alpha i},\quad
\bar\lambda_{\dot\alpha}^{\prime\, i} =e^{-i(b-ic)}\bar\lambda_{\dot\alpha}^i ,\quad
e^{\prime\, a}_m =e^{2c}e^{a}_m\,.
$$
We can fix the real parameters $b$, $c$ by imposing of the constraints
\begin{equation}\label{nor2}
A\equiv\lambda^{\alpha i} \lambda_{\alpha i}-T=0 \,, \qquad \bar
A\equiv\bar\lambda_{\dot\alpha}^{i} \bar\lambda^{\dot\alpha}_i - T =0\,.
\end{equation}
If we introduce the variables $v_{\alpha i} ={\textstyle\sqrt{\frac{2}{T}}}\,
\lambda_{\alpha i}$, $\bar v_{\dot\alpha}^i = {\textstyle\sqrt{\frac{2}{T}}}\,
\bar\lambda_{\dot\alpha}^i$ we get the orthonormality relations for the spinorial Lorentz
harmonics~\cite{BZ}.

If we impose the constraints~(\ref{nor2}) the model~(\ref{SSTV}) can be rewritten in the
following way
\begin{equation}\label{act-harm}
S=\int d^2\xi \left[e e^m_a \tilde{\lambda}_{\dot\alpha}^i (\rho^a)_i{}^j\lambda_{\alpha
j}\, \partial_m X^{\dot\alpha\alpha}+ {\textstyle \frac{T}{2}}\,e + \Lambda A +
\bar\Lambda\bar A\right]
\end{equation}
where the spinors $\lambda$, $\bar\lambda$ are constrained by the relations~(\ref{nor2}),
which are imposed additionally in~(\ref{act-harm}) by the Lagrange multipliers. It is easy
to see that introducing the light cone notations on the world sheet for the zweibein
$e_m^{++} = e_m^{0}+e_m^{1}$, $e_m^{--} = e_m^{0}-e_m^{1}$ and following  Weyl
representation for Dirac matrices in two dimensions we obtain harmonic string
action~\cite{BZ,BAM,Uv}.

{\bf Mixed formulations with tensorial momenta.} The zweibein $e_m^a$ can be expressed from
the action~(\ref{SSTV}) as follows
\begin{equation} \label{resol-tens}
e_m^a ={\textstyle\frac{2T}{(\lambda\lambda)(\bar\lambda\bar\lambda)}}\,\,
\,\tilde{\lambda}_{\dot\alpha}^i (\rho^a)_i{}^j\lambda_{\alpha j}\,\, \partial_m
X^{\dot\alpha\alpha}
\end{equation}
Substitution of the relation~(\ref{resol-tens}) in the
action~(\ref{action-tw}) provides the following string action
\begin{equation}\label{tens-3}
S\!=\!\!{\textstyle\sqrt{2}}\!\!\int \!\!d^2\xi\epsilon^{mn}\!\! \left(\! P_{\alpha\beta}\,
\partial_m X^{\dot\gamma\alpha} \partial_n X_{\dot\gamma}^{\beta}\!+\!
\bar P_{\dot\alpha\dot\beta} \,\partial_m X^{\dot\alpha\gamma} \partial_n
X^{\dot\beta}_{\gamma}\! \right)
\end{equation}
where the composite second rank spinors
\begin{equation}\label{tens-mom}
P_{\alpha\beta}={\textstyle\frac{\sqrt{2}T}{({\lambda}{\lambda})}}\,
\lambda_{(\alpha}^1\lambda_{\beta)}^2\,, \qquad \bar
P_{\dot\alpha\dot\beta}={\textstyle\frac{\sqrt{2}T}{(\bar{\lambda}\bar{\lambda})}}\,
\bar{\lambda}_{(\dot\alpha}^1 \bar{\lambda}_{\dot\beta)}^2\,.
\end{equation}
satisfy the constraints
\begin{equation}\label{tens-mom-con}
P^{\alpha\beta}P_{\alpha\beta}=- {\textstyle\frac{T^2}{4}}\,, \qquad \bar
P^{\dot\alpha\dot\beta}\bar P_{\dot\alpha\dot\beta}=-{\textstyle\frac{T^2}{4}}\,.
\end{equation}
Using fourvector notation
$P_{\alpha\beta}=P_{\mu\nu}\sigma^{\mu\nu}_{\alpha\beta}$, $\bar
P_{\dot\alpha\dot\beta}=-P_{\mu\nu}
\sigma^{\mu\nu}_{\dot\alpha\dot\beta}$ the
relations~(\ref{tens-mom-con}) take the form
\begin{equation}\label{tens-mom-con1}
P^{\mu\nu}P_{\mu\nu} = -{\textstyle\frac{T^2}{4}}\,, \qquad
P^{\mu\nu}\tilde P_{\mu\nu} =0\,.
\end{equation}

Using additionally the conditions~(\ref{nor2}) the formulation~(\ref{tens-3}) produces
harmonic string action with tensorial string momenta (see also~\cite{GZ}).

\section{Purely twistorial formulation}

Let us introduce second half of twistor coordinates $\mu^{\dot\alpha}_i$,
$\bar{\mu}^{\alpha i}$ by employing Penrose incidence relations generalized for string
\begin{equation}\label{Pen-inc}
\mu^{\dot\alpha}_i=X^{\dot\alpha\alpha}\lambda_{\alpha i} \,,\qquad \bar{\mu}^{\alpha
i}=\bar{\lambda}_{\dot\alpha}^i X^{\dot\alpha\alpha}\,.
\end{equation}
Incidence relations~(\ref{Pen-inc}) with real space--time string
position $X^{\dot\alpha\alpha}$ imply that the twistor variables
satisfy the constraints
\begin{equation}\label{V}
V_i{}^j \equiv \lambda_{\alpha i}\bar{\mu}^{\alpha j} - \mu^{\dot\alpha}_i
\bar{\lambda}_{\dot\alpha}^j \approx 0
\end{equation}
which are antiHermitian ($(\overline{V_i{}^j})=-V_j{}^i$).

Let us insert the relations~(\ref{Pen-inc}) into (\ref{act-harm}). Using
$$
P_{\alpha\dot\alpha}^{\,m}\partial_m X^{\dot\alpha\alpha}= {\textstyle \frac{1}{2}}e e_a^m
\left[\tilde{\lambda}_{\dot\alpha}\rho^a\partial_m \mu^{\dot\alpha}
-\tilde{\mu}^\alpha\rho^a\partial_m\lambda_\alpha \right] + {\it c.c.}
$$
we obtain the first order string action in twistor formulation
\begin{eqnarray}
S \!\!\!\!\!\! &&= \int d^2\xi   \left\{ {\textstyle \frac{1}{2}} e e_a^m
 \left[\tilde{\lambda}_{\dot\alpha}\rho^a\partial_m
\mu^{\dot\alpha} -\tilde{\mu}^\alpha\rho^a\partial_m\lambda_\alpha + {\it c.c.}\right]\right.  + \nonumber\\
&& \qquad\qquad\quad \left. + {\textstyle \frac{T}{2}}\,e + \Lambda_j{}^i V_i{}^j  +
\Lambda A + \bar\Lambda\bar A\right\} \label{action-tw}
\end{eqnarray}
where $\Lambda_i{}^j=-(\overline{\Lambda_j{}^i})$, $\Lambda$,
$\bar\Lambda$ are the Lagrange multipliers.

Introducing the string twistors
$$
Z_{Ai}=(\lambda_{\alpha i}, \mu_i^{\dot\alpha}), \,\,\,\, \bar
Z^{Ai}=(\bar\mu^{\alpha i}, -\bar\lambda_{\dot\alpha}^{i} ),
\,\,\,\, \tilde Z^{Ai}=\bar Z^{Aj}(\rho^0)_j{}^i,
$$
the constraints~(\ref{V}) are rewritten as
\begin{equation}\label{V-Z}
V_i{}^j = Z_{Ai}\bar Z^{Aj}\approx 0\,.
\end{equation}
Substituting the equations of motion for zweibein $e_m^a$
\begin{equation}\label{e-am}
e_m^a =-{\textstyle\frac{1}{T}} \left[\partial_m\tilde Z^{A}\rho^a
Z_{A} - \tilde Z^{A}\rho^a \partial_m Z_{A} \right]
\end{equation}
in the action~(\ref{action-tw}) we obtain our basic twistorial
string action~\cite{FL}:
\begin{eqnarray}
&& \qquad\qquad\qquad\qquad\qquad S = \int d\,^2\xi\,{\cal L}\,, \label{S2}\\
{\cal L} \!\!&\!\! = \!\!&\!\! {\textstyle \frac{1}{4T}} \epsilon^{mn}\epsilon_{ab}
\left[\partial_m\tilde Z^{B}\rho^a Z_{B} - \tilde Z^{B}\rho^a
\partial_m Z_{B} \right]\left[\partial_n\tilde Z^{A}\rho^b Z_{A} - \tilde Z^{A}\rho^b
\partial_n Z_{A}
\right]+  \nonumber\\
&& +\Lambda_j{}^i V_i{}^j  + \Lambda A + \bar\Lambda\bar A \,. \label{action-tw-2}
\end{eqnarray}

Using explicit form of $D=2$ Dirac matrices we obtain that the first term in the
Lagrangian~(\ref{action-tw-2}) equals to
\begin{equation}\label{forma}
 {\textstyle \frac{1}{T}}\epsilon^{mn}
\!\!\left[\partial_m\bar Z^{A1}\! Z_{A1} \!-\! \bar Z^{A1} \partial_m Z_{A1}
\right]\!\left[\partial_n\bar Z^{B2}\! Z_{B2} \!-\! \bar Z^{B2} \partial_n Z_{B2}\right]
\end{equation}
i.e. the action~(\ref{action-tw-2}) is induced on the world--sheet by the canonical
2--form~(\ref{2-form}) with supplemented constraints~(\ref{nor2}) and~(\ref{V}).

\section{Conclusions}
In this report we presented the classical
 master twistor formulation of tensionful string~(\ref{S2}) and
its links with various bosonic string models. These links can be represented by the
following diagram:
\\
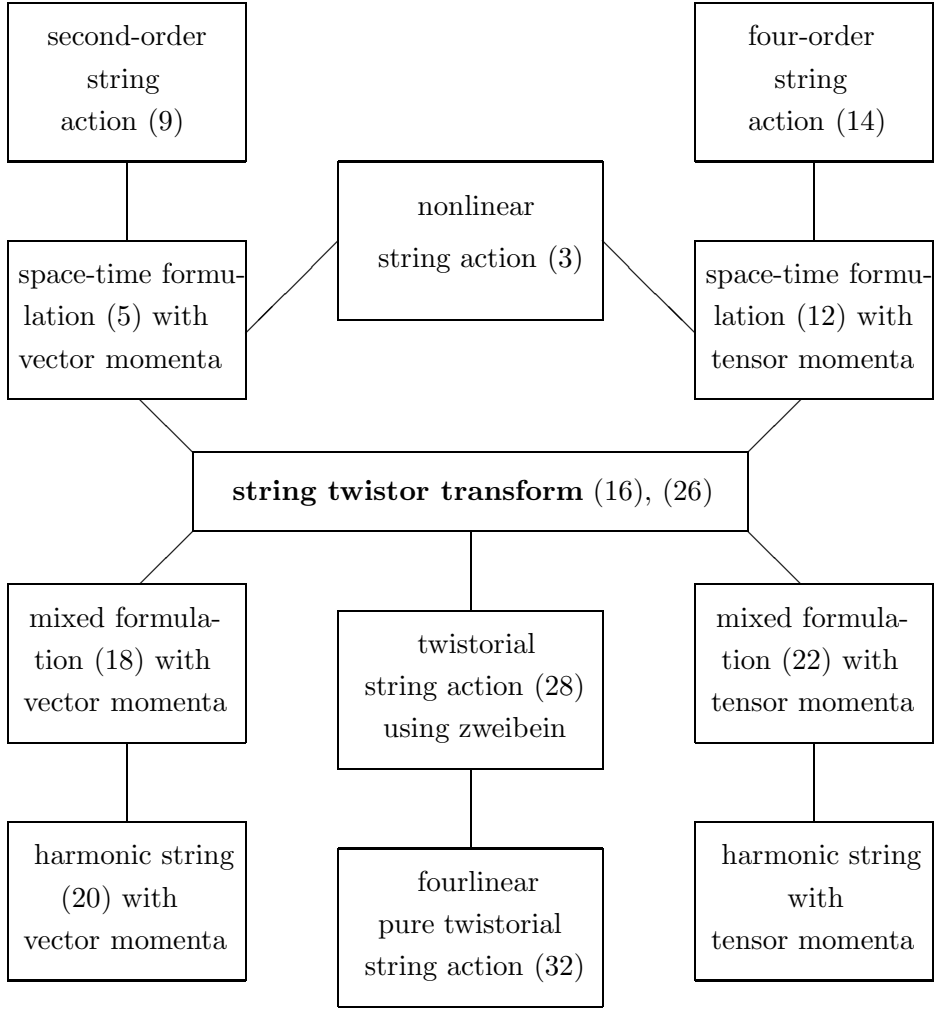
\begin{picture}(350,395)
\put(125,0){\line(1,0){100}} \put(125,0){\line(0,1){60}} \put(225,0){\line(0,1){60}}
\put(125,60){\line(1,0){100}} \put(135,12){string action (\ref{action-tw-2})}
\put(140,28){pure twistorial} \put(155,44){fourlinear}

\put(125,90){\line(1,0){100}} \put(125,90){\line(0,1){60}}
\put(225,90){\line(0,1){60}}\put(125,150){\line(1,0){100}} \put(142,102){using zweibein}
\put(135,118){string action (\ref{action-tw})} \put(155,134){twistorial}

\put(70,180){\line(1,0){210}} \put(70,180){\line(0,1){30}} \put(280,180){\line(0,1){30}}
\put(70,210){\line(1,0){210}} \put(85,192){{\bf string twistor transform} (\ref{P-res-st}),
(\ref{Pen-inc})}

\put(0,10){\line(1,0){90}} \put(0,10){\line(0,1){60}} \put(90,10){\line(0,1){60}}
\put(0,70){\line(1,0){90}} \put(6,22){vector momenta} \put(20,38){(\ref{act-harm}) with}
\put(10,54){harmonic string}

\put(0,100){\line(1,0){90}} \put(0,100){\line(0,1){60}}
\put(90,100){\line(0,1){60}}\put(0,160){\line(1,0){90}} \put(6,112){vector momenta}
\put(10,128){tion (\ref{SSTV}) with} \put(8,144){mixed formula-}

\put(260,10){\line(1,0){90}} \put(260,10){\line(0,1){60}} \put(350,10){\line(0,1){60}}
\put(260,70){\line(1,0){90}} \put(266,22){tensor momenta} \put(295,38){with}
\put(270,54){harmonic string}

\put(260,100){\line(1,0){90}} \put(260,100){\line(0,1){60}}
\put(350,100){\line(0,1){60}}\put(260,160){\line(1,0){90}} \put(266,112){tensor momenta}
\put(270,128){tion (\ref{tens-3}) with} \put(268,144){mixed formula-}

\put(0,230){\line(1,0){90}} \put(0,230){\line(0,1){60}} \put(90,230){\line(0,1){60}}
\put(0,290){\line(1,0){90}} \put(4,242){vector momenta} \put(6,258){lation (\ref{act-s})
with} \put(4,274){space-time formu-}

\put(0,320){\line(1,0){90}} \put(0,320){\line(0,1){60}}
\put(90,320){\line(0,1){60}}\put(0,380){\line(1,0){90}} \put(20,332){action
(\ref{action-2})} \put(30,348){string} \put(15,364){second-order}

\put(260,230){\line(1,0){90}} \put(260,230){\line(0,1){60}} \put(350,230){\line(0,1){60}}
\put(260,290){\line(1,0){90}} \put(266,242){tensor momenta} \put(267,258){lation
(\ref{tens-4}) with} \put(264,274){space-time formu-}

\put(260,320){\line(1,0){90}} \put(260,320){\line(0,1){60}}
\put(350,320){\line(0,1){60}}\put(260,380){\line(1,0){90}} \put(280,332){action
(\ref{tens-5})} \put(290,348){string} \put(280,364){four-order}

\put(125,260){\line(1,0){100}} \put(125,260){\line(0,1){60}} \put(225,260){\line(0,1){60}}
\put(125,320){\line(1,0){100}} \put(140,280){string action (\ref{action-NG})}
\put(155,300){nonlinear}

\put(175,60){\line(0,1){30}} \put(175,150){\line(0,1){30}}

\put(45,70){\line(0,1){30}} \put(305,70){\line(0,1){30}}

\put(70,180){\line(-1,-1){20}} \put(280,180){\line(1,-1){20}}

\put(70,210){\line(-1,1){20}} \put(280,210){\line(1,1){20}}

\put(45,290){\line(0,1){30}} \put(305,290){\line(0,1){30}}

\put(125,290){\line(-1,-1){35}} \put(225,290){\line(1,-1){35}}

\end{picture}

There are several ways to extend the studies presented briefly in this report which are now
under consideration:

i) One can consider the Hamiltonian formulation of the twistorial string
 action
(\ref{action-tw-2}),
   with first and second class constraints.
   One should observe a convenient property of the fourlinear action
(\ref{action-tw-2})  -- it is linear in time derivative, in analogy to the
    momentum formulations of the superparticle models. The aim of our studies
    is to obtain the Gupta-Bleuler type of quantization of the model
(\ref{action-tw-2}).

    ii) One can show that the PB of the constraints
    (\ref{V-Z})
     describe $U(2)$ algebra.
    Following the discussion of massive spinning particle model
    in $2$-twistor space
    \cite{azfrlu}--\cite{FFLM}.
    one can interpret  the four $V_{i}^{\, j}$ generators as
    describing local density of spin and electric charge of the string.

    iii) The presented links between various bosonic string models can be
    extended supersymmetrically in two-supertwistor space, by applying
    already developed supersymmetrization techniques \cite{SSTV,BZ}.

 iv) In the approach of Witten and Berkovits
\cite{Wit,BerWit}
   the twistor string model
 is endowed with single (super)twistor target space, and the correspondence
 with Penrose theory is obtained only on the level of quantized
 twistorial string.
 Our twistorial model is classically equivalent to the standard bosonic string
  model.
 If the quantization of our model can be achieved, it will be possible
  to relate our and Witten`s approach, which we expect will strictly  overlap only
  for very restricted class of string modes.

\subsection*{Acknowledgments}

 One of us (S.F.) would like to thank the Organizers of {\it the 22nd
Max Born Symposium} for a very pleasant atmosphere. We would like to thank E.~Ivanov and
D.~Sorokin for valuable remarks. The work of J.L. was supported by KBN grant 1 P03B 01828.
The work of S.F. was supported in part by the RFBR grant 06-02-16684, the grant
INTAS-05-7928 and the grants from Bogoliubov--Infeld and Heisenberg-Landau programs.

\end{document}